\documentclass[intlimits,twoside,a4paper]{article}
\usepackage{graphics}
\usepackage{amsmath}
\usepackage{color}

\usepackage[T2A]{fontenc}
\usepackage[cp1251]{inputenc}

\usepackage[eqsecnum]{cmpj2}

\issue{2016}{19}{1}{13002}
\doinumber{10.5488/CMP.19.13002}

\title[Critical point calculation for binary mixtures of symmetric non-additive hard disks]
{Critical point calculation for binary mixtures \\ of symmetric non-additive hard disks}

\author[W.T. G\'o\'zd\'z, A. Ciach]{W.T. G\'o\'zd\'z, A. Ciach}

\address{Institute of Physical Chemistry Polish Academy of Sciences, Kasprzaka 44/52, 01-224 Warsaw, Poland}

\date{Received November 6, 2015, in final form November 29, 2015}
\authorcopyright{W.T. G\'o\'zd\'z, A. Ciach, 2016}

\begin{document}

\maketitle
\begin{abstract}
We have calculated the values of critical packing fractions for the mixtures of symmetric non-additive
hard disks. An interesting feature of the model is the fact that the internal energy
is zero and the phase transitions are entropically driven. A cluster algorithm for Monte Carlo simulations
in a semigrand ensemble was used. The finite size scaling analysis was employed to compute the
critical packing fractions for infinite systems with high accuracy for a  range of non-additivity parameters
wider than in the previous studies.
\keywords phase coexistence, critical point, finite size scaling, Monte Carlo simulations
\pacs 05.10.Ln, 05.20.Jj, 05.70.Jk
\end{abstract}
%
%
%
\section{Introduction}

Two-dimensional fluid mixtures are quite common in soft-matter and in biological systems. Important examples are particles
adsorbed at interfaces, on surfaces of  pores in porous materials and biological membranes. When
 the adsorbed fluid forms a monolayer, it may be
modelled as  a two-dimensional system.
The  phenomenon of adsorption has been intensively studied, and one of the main contributors to the field
is Stefan Sokolowski~\cite{gozdz:91:0,bucior:03:0,duda:04:0,patrykiejew:07:0,borowko:12:0,patrykiejew:12:0}.
An interesting question that is not fully solved yet is the phase separation in a binary mixture on surfaces with different curvatures.
This question is important not only for the adsorption on curved surfaces present in porous materials, but also for the properties of
biological membranes surrounding organelle. In living organisms, the membranes are close to the critical point of the demixing
transition \cite{veatch:07:0,veatch:09:0}. Therefore,
the phase behavior of multicomponent two dimensional  fluids and, particularly, the critical behavior
may be of biological importance. During the phase transition, a small change of thermodynamic parameters
causes a large change of the composition. It may result in significant shape transformations
of the biological membranes since their shape is linked with their composition~\cite{gozdz:12:0}.
Especially interesting are the membranes which form triply periodic bilayers~\cite{gozdz:96:1,dotera:12:0}, as well as  porous materials
 whose internal surfaces are similar to minimal surfaces (the average curvature vanishes at every point, and the Gaussian curvature is negative).

The phase separation in binary fluid mixtures belongs to the Ising universality class, and
the universal properties of the two-dimensional Ising model are well known from exact results~\cite{onsager:44:0}.
The nonuniversal properties, however, should be determined for each particular system, and the only
feasible method for a surface with arbitrary curvature is by computer simulations. Thus,
it is important to develop a simulation procedure that is fast, efficient and accurate.
Moreover, it is important to very accurately determine the critical parameters of a generic model on a flat surface, so that
the results of approximate theories or the results of simulations on curved surfaces could be compared with reliable data.
The lattice model
is not appropriate for investigations of the properties of curved surfaces. Therefore, in this work we have chosen a generic continuous model
of non-additive hard core mixtures.
Some real phenomena, which can be
modelled by a mixture of non-additive hard disks, are discussed,
for example, in reference \cite{fiumara:14:0} and in references cited herein.
The behavior of the hard core mixtures has been studied in
bulk~\cite{amar:89:0,lomba:96:0,johnson:97:0,saija:98:0,gozdz:01:1,gozdz:03:0,buhot:05:0,pelicane:14:0} and
in restricted geometry~\cite{duda:04:0,gozdz:05:1,lucentini:08:0,almarza:15:0} in three dimensional systems.
Much less attention was paid to  the two dimensional
systems~\cite{almarza:15:0,buhot:05:0,fiumara:14:0,munoz-salazar:10:0,nielaba:96:0,saija:02:0}.

We study the mixture of symmetric non-additive hard disks with the interaction potential defined by:
%
%
 \begin{equation}
 U_{\alpha\gamma}(r)= \left\{
 \begin{array}{lll}
 \infty &\;\; \mbox{if}  & \;\;\; \hbox{$r < \sigma_{\alpha\gamma}$}, \\
 0      &\;\; \mbox{if}  & \;\;\; \hbox{$r > \sigma_{\alpha\gamma}$},
 \end{array}
 \right.
 \label{e:intermolecular_potential}
 \end{equation}
 where $r$ is the distance between the centers of two disks, indexes  $\alpha \in \{A,B\} $ and $\gamma \in \{A,B\} $
 describe the  species. The length scale is set by the $A$-component hard-disks diameter, $\sigma_{AA}=1$.
 For symmetric non-additive mixtures,
%
%
 \begin{equation}
 \sigma_{BB}=\sigma_{AA}
 \end{equation}
and
%
%
 \begin{equation}
 \sigma_{AB}=\frac{1}{2}(\sigma_{AA} + \sigma_{BB}) (1+\Delta),
 \end{equation}
 where $\Delta$ is the non-additivity parameter.   We study the mixtures of positive non-additivity with different
 values of the parameter $\Delta$. This potential is an idealization of interactions in
 a  mixture of identical colloid particles with surfaces covered with polymeric brushes
 of two  types, $A$ and $B$. The polymeric brushes of different type effectively repel each other,
 but the polymers of the same type can interpenetrate. For this reason, the separation between the  like particles can be shorter than between different particles.

Quite  evidently,  this is an athermal model, because all the allowed configurations
 are of the same energy.  Nevertheless,  such mixtures are capable of separating into two phases, i.e., one phase rich in component  $A$ and the other one rich in component $B$. The phase separation is not induced by competition between the internal energy
 and the entropy as in standard systems, but rather by competition between the entropy of mixing and the entropy
 associated with the area available for the particles.
 An increase of the packing  fraction
  defined as
%
%
 \begin{equation}
  \eta= \eta_A +\eta_B =\frac{\pi}{4} \frac{N_A\sigma_{AA}}{S} + \frac{\pi}{4} \frac{N_B\sigma_{BB}}{S},
  \end{equation}
where $S$ is the surface area of the system and $N_A$ and $N_B$ are the numbers of the $A$ and $B$ particles,
 leads to a larger decrease of the available area in a homogeneous mixture than in a phase-separated system.
 This effect starts to dominate over
 the decrease of the entropy of mixing in a phase-separated system for $\eta>\eta_\text{c}$. Thus, $\eta_\text{c}$ plays the role analogous
 to the critical temperature $T_\text{c}$, and $(\eta_\text{c}-\eta)/\eta_\text{c}$ plays a role analogous
 to $(T-T_\text{c})/T_\text{c}$ in standard mixtures with
 interacting particles. Simulation results \cite{gozdz:03:0} show that this model system belongs to the Ising universality class.
 The universal properties of the model are known from the exact solution of the Ising model in two dimensions, but the
 nonuniversal properties, such as $\eta_\text{c}$, should be obtained by simulations.

 At the absence of an external field, the symmetry of interactions implies, for the two coexisting
 phases I and II, the following relations:
%
%
 \begin{equation}
   x_A^\text{I}=x_B^\text{II}, \hskip 1cm
   x_B^\text{I}=x_A^\text{II},
 \end{equation}
and
%
%
 \begin{equation}
   \mu_A^\text{I}=\mu_A^\text{II}=\mu_B^\text{I}=\mu_B^\text{II},
 \end{equation}
where $\mu_\alpha^\text{I}$ and $x_\alpha^\text{I}$ are the chemical potential and the composition of the  component $\alpha$ in the $i$-th
phase, respectively~\cite{gozdz:95:0,gozdz:03:0}. Here, the difference of the chemical potentials $h=\mu_A-\mu_B$ plays
the role of an external field.

We are interested in the critical properties of a mixture when the external field is zero. Along the symmetry line $h=0$,
the composition of the coexisting phases is symmetric, and therefore the critical point is at the  concentration $x_\text{c}=N_A/(N_A+N_B)=0.5$.

\section{The method}

 We model an open system in contact with a reservoir by using the
Semigrand~\cite{kofke:88:0,gozdz:03:0} Monte Carlo~\cite{metropolis:49:0,metropolis:54:0} simulation method.
Using this method, the system is simulated under a constant total number of particles $N$, total volume $V$, temperature $T$, and
the difference of the chemical potential of one species with respect to an arbitrarily chosen species $\Delta\mu$.
Thus, the number of molecules of each species fluctuates, while the total density remains constant.
The semigrand ensemble is superior to
the grand ensemble in simulating dense fluids because the particle insertion moves are inefficient for dense fluids.
 For a symmetric binary hard
disks mixture, the internal energy and the chemical potential difference are both zero.

The realization of the semigrand ensemble Monte Carlo simulation for symmetric non-additive hard disks requires two kinds of moves:
translation and identity change.
The identity change moves can be implemented in the following  way. A molecule  can be chosen randomly from all the molecules,
and the identity change move is always accepted if there is no overlap between the particles
after the identity change. Such a procedure works
quite well for a small  number of molecules, up to a few hundred. The  simulations near the critical point might be, however,
time consuming. Therefore, we use a cluster algorithm to perform the simulations for larger system
sizes~\cite{johnson:97:0,almarza:15:0}. The idea of the cluster moves is as follows.
The system is divided into a set of clusters. The molecules belong to a cluster if the distance from any molecule
to any other molecule is smaller than $\sigma_{AA}+\Delta$. When the clusters are
formed, the identity of all molecules in each cluster is randomly changed with the probability $p=0.5$.
We identify the clusters using the SLINK algorithm~\cite{florek:51:0,florek:51:1,sibson:73:0}, which is fast and does not require a huge amount of computer
memory. We have performed the calculations using either local MC moves or cluster moves. The results obtained by both
methods were consistent, but the time of calculations was much shorter for the cluster algorithm.

For the translation moves, the maximum displacement is chosen to obtain $50\%$ acceptance ratio.
The identity-change and the translation moves are chosen randomly with the ratio of $N$ translations per
one cluster identity change move. We have performed calculations with a square or with a rectangular simulation box.
The aspect ratio of the rectangular box is taken as $\sqrt{3}:2$ to allow for the arrangement of the molecules
into a triangular lattice.  Periodic boundary conditions are employed. We have not observed any dependence of
the results of calculations on
the shape of the simulation box when the simulations were performed for fluid mixtures. The averages are taken over $10^6$ Monte Carlo cycles, where a cycle consists of $N$ translation and one cluster identity change move.

In molecular simulations, the results of calculations depend on a system size.  The dependence is more pronounced for
the calculations near the critical point since the correlation length becomes larger and larger when the system is
closer and closer to the critical point. When the correlation length is larger than the size of the computational box,
the results of the calculations become biased.  That is why it may
be very difficult to perform exact calculations of the critical point parameters.

 When the system is in the two phase region, we do not
always have only one phase in the simulation box  during the simulation
 in the semigrand Monte Carlo method. It is possible that in the simulation box we will have either the first or the second
phase. With some frequency, the first phase disappears and the second phase appears  and vice versa. The higher is the
frequency of this change, the closer  the system is to the critical point. That is why it is hard
to calculate the concentration at the coexistence as an ensemble average. One may try to overcome this problem by taking
the most probable value of the concentration instead of the ensemble average, but near the critical point, this procedure
may be problematic especially for two dimensional fluids. The shape of the coexistence curve for the two dimensional
fluid is much flatter than for the three dimensional fluids.  The critical exponent for the two dimensional fluids is
$\beta=1/8$ while for the three dimensional fluids, it is $\beta\approx 0.3258$. The distribution of the concentration is
not sharply peaked and it is difficult to obtain the most probable value of the concentration with a sufficiently high accuracy.

Fortunately, it is possible to use the calculations performed for the systems of finite size to get the data
on the infinite systems by using the concept of finite-size  scaling \cite{fisher:72:0,barber:83:0,wilding:95:0,gozdz:03:0}.
For each value of the non-additivity
parameter $\Delta$, the
critical packing fraction of the
infinite system, $\eta_\text{c}(\infty)$, can be calculated
from the set of apparent critical packing fractions in systems with $N$ particles, $\eta_\text{c}^*(N)$,
according to the relation \cite{barber:83:0,wilding:95:0,gozdz:03:0}:
\begin{equation}
\eta_\text{c}^*(N)- \eta_\text{c}(\infty) \propto  N^{-1/(d\nu)},
\label{e:scaling}
\end{equation}
where the critical exponent $\nu$ is equal to 1 for the 2D Ising universality class, and   $d=2$ for the two dimensional systems.
The apparent critical point in the finite-size system can be determined from the
  distribution of the order parameter, $P_N(m)$ calculated in the simulations
as a histogram.
 The order parameter in the case of the non-additive hard disks system is the concentration, $m=x-x_\text{c}$, where
$x=N_{A}/(N_{A}+N_{B})$ and $x_\text{c}$ is the critical concentration which is exactly $1/2$ for symmetric mixtures.
The distribution $P_N(m)$ is rescaled in such a way as to have a unit norm and a unit variance, and the rescaled distribution
is denoted by $P^*_N(y)$, where the rescaled order parameter is $y=a_m^{-1}N^{\beta/(d\nu)}m$ with $a_m$ denoting
a proportionality constant.  For the apparent
critical packing fraction  $\eta_\text{c}^*(N)$,
  $P^*_N(y)$ has an universal shape  $P^*(y)$ \cite{wilding:95:0,gozdz:03:0}.
Since this model belongs to the Ising universality class, the universal function   $P^*(y)$
is known~\cite{nicolaides:88:0,liu:10:0}.

 \section{Results}

\begin{figure}[!h]
\begin{center}
\includegraphics[width=0.50\textwidth, angle = 0 ]{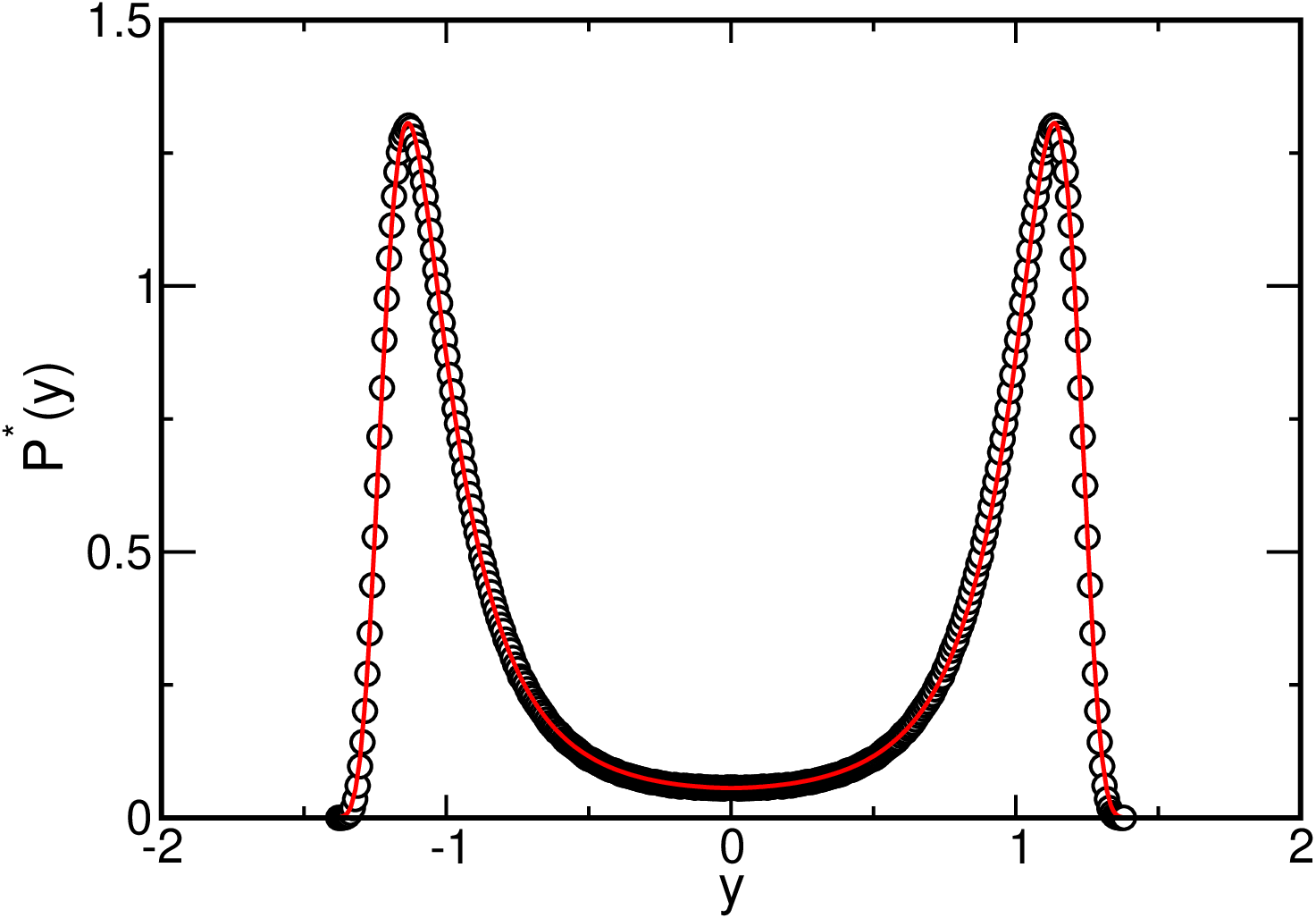}
\end{center}
\vspace{-5mm}
  \caption{(Color online) The normalized distribution of the order parameter $P^*_N(y)$
  (open circles) for $\Delta=0.2$ and $\sqrt{N}=18$, expressed as the function of
  the scaling variable $y=a_m^{-1}N^{\beta/(d\nu)}m$, at the packing
  fraction $\eta=0.5384$.  The solid line is the universal function $P^*(y)$
  for the Ising model \cite{liu:10:0}. The critical exponent $\nu$ is equal to 1, and $a_m$ is  a proportionality constant. }
  \label{f:order_parameter}
\end{figure}

\begin{figure}[!t]
  \begin{center}
\includegraphics[width=0.51\textwidth,angle=0]{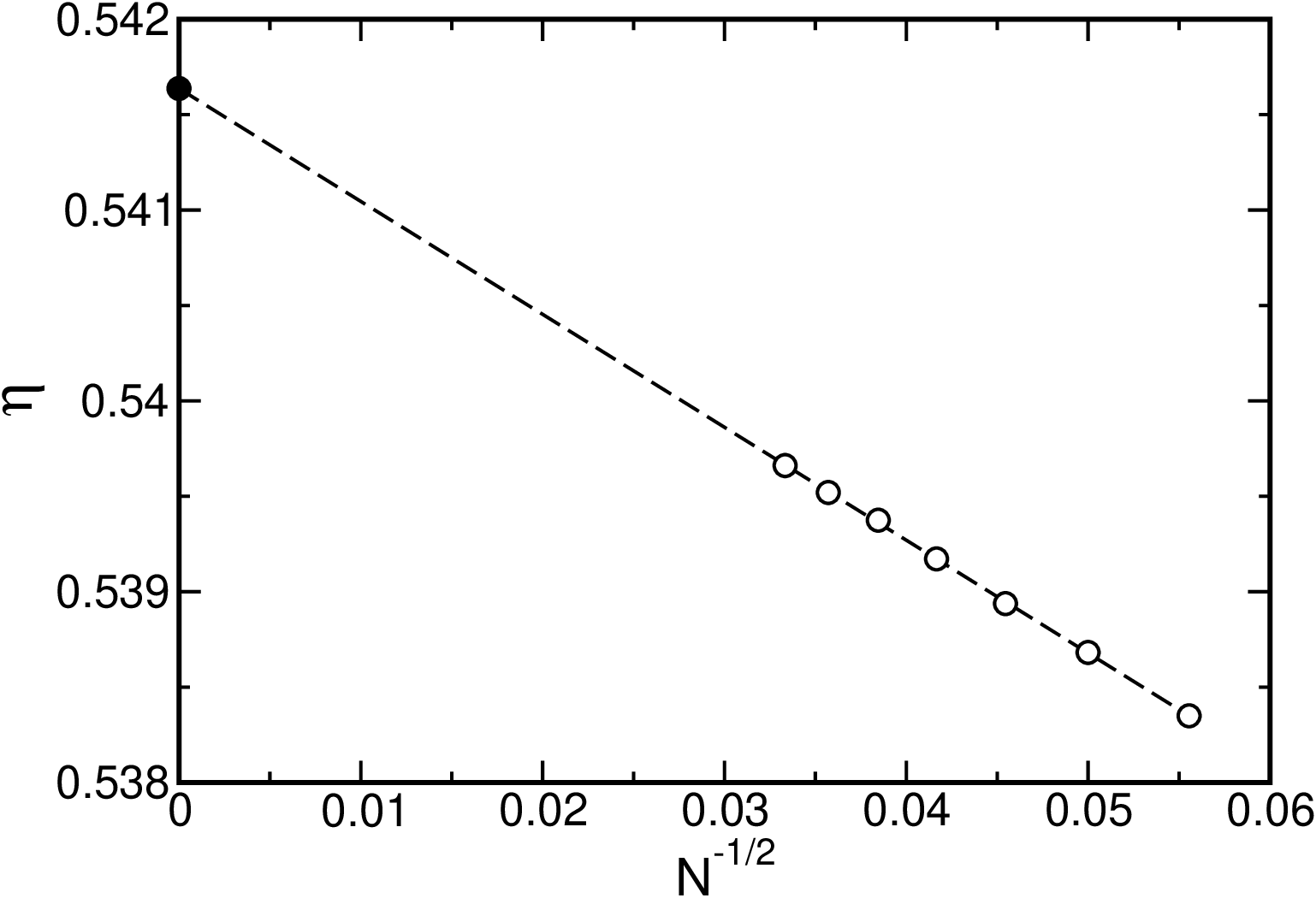}
  \end{center}
  \vspace{-5mm}
  \caption{Scaling of the apparent critical packing fraction with the system size $\sqrt{N}=\{18,20, 22, 24, 26, 28, 30\}$
  for $\Delta=0.2$. The open circles represent  the apparent critical packing fractions for different system sizes
 obtained by matching $P^*_N(y)$ with the universal function as in figure~\ref{f:order_parameter}.
  The solid circle is the critical packing fraction for the infinite system [see equation (\ref{e:scaling})].
  The dashed line is the least square fit
  of the apparent critical packing fractions to a straight line.  }
  \label{f:scaling}
\end{figure}

 The order-parameter distribution  $P_N(m)$ is calculated in the simulations as a histogram, with the number
  of bins equal to the number
of particles $N$. The function $P_N(m)$ obtained in simulations
for fixed $N$ and various $\eta$ is then rescaled in such a way as to have a unit norm and  a unit variance.
The rescaled function $P^*_N(y)$ is compared with  $P^*(y)$ for several values of $\eta$,
and the best fit gives us $\eta_\text{c}^*(N)$.
Figure~\ref{f:order_parameter} shows the order-parameter distribution function $P^*_N(y)$ calculated in the
simulations for the system with $\Delta=0.2$ and $N=324$ disks,
compared with the distribution function for the 2D Ising model~\cite{liu:10:0}.
 $P^*_N(y)$ agrees very well with the universal distribution $P^*(y)$ for the value of $\eta$ that we identify
 with the apparent critical volume fraction  $\eta_\text{c}^*(N)$.
In the same way, we obtain apparent critical packing fractions for a set of systems with a different size.
In figure~\ref{f:scaling},
we present the plot of apparent critical densities for different system sizes calculated for the
non-additivity parameter $\Delta=0.2$.
The apparent critical packing fractions are estimated for a set of finite systems with $\sqrt{N}=\{18,20, 22, 24, 26, 28, 30 \}$.
The critical packing fraction as a function of $N^{-1/2}$ for an infinite system was obtained by fitting the set of apparent critical
packing fractions to a straight line and extrapolating the value of the critical packing fraction at infinity.
The same procedure was used to determine the critical packing fractions for all the values of the non-additivity parameter $\Delta$.

In practice, the shape of the rescaled distribution function $P^*_N(y)$ is compared with the universal distribution
of the order parameter, $P^{*}(y)$,  for the first estimation of the apparent critical packing fraction.
In order to determine the precise value of the apparent critical packing fraction, the fourth order cumulants,
\begin{equation}
U_N=1-\frac{\langle m^4 \rangle}{3\langle m^2 \rangle ^2}\,.
\label{e:binder_cumulant}
\end{equation}
are calculated for the values of the packing fraction $\eta$ for which the distribution functions are the most similar
to the universal distribution $P^*(y)$. To calculate the matching point, the values of the cumulant $U_N(\eta)$
have been  interpolated near the universal value $U_N^*$. We use the fourth order
cumulant $U_N$ \cite{binder:81:0}, because its value at a fixed point has been already calculated for the 2D Ising
universality class.

\begin{figure}[!b]
  \begin{center}
\includegraphics[width=0.50\textwidth,angle=0]{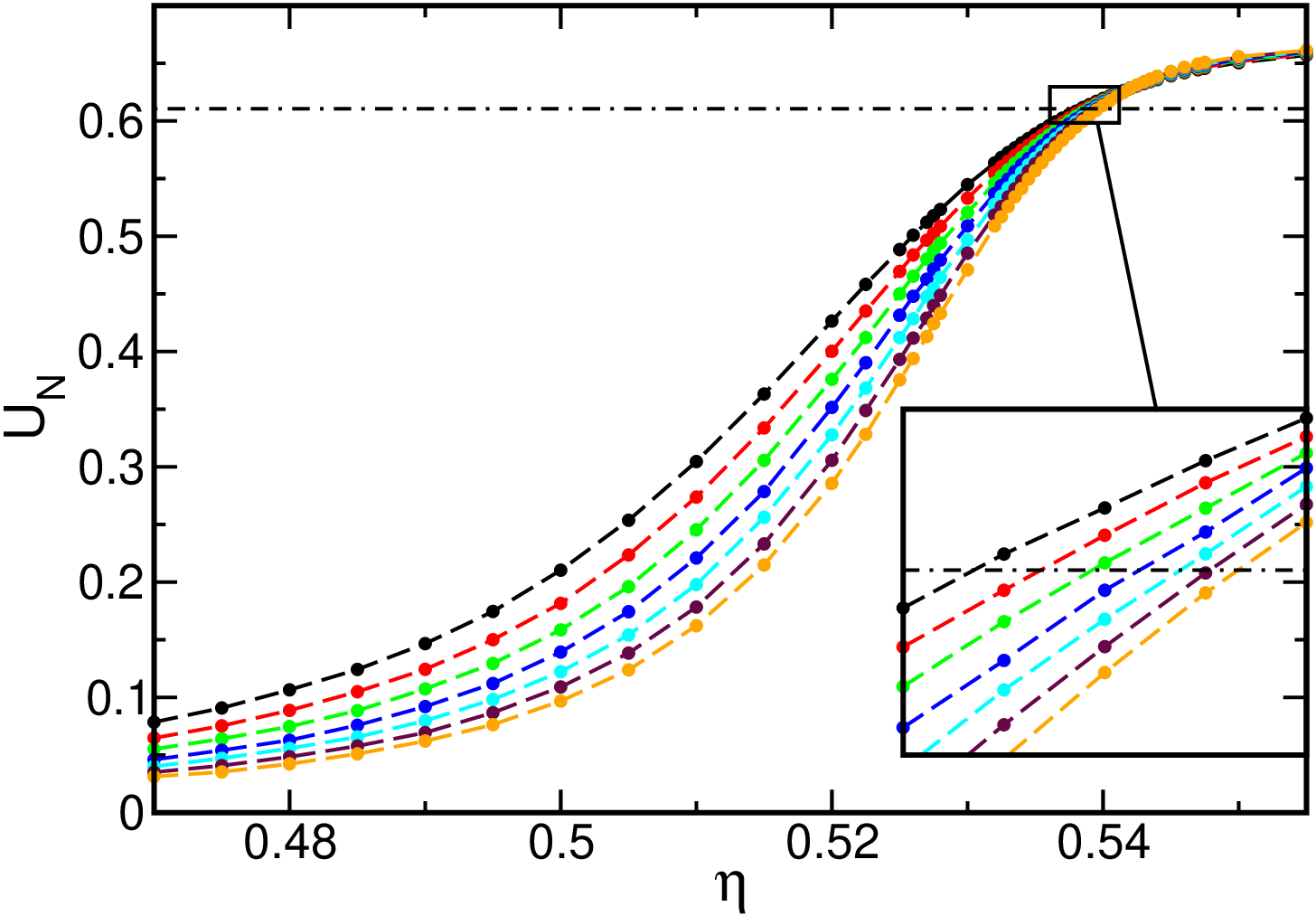}
  \end{center}
  \vspace{-5mm}
  \caption{(Color online) Fourth order cumulant, $U_N(\eta)$,  as a function of the packing fraction $\eta$ for
  the non-additivity parameter $\Delta=0.2$ and
  $\sqrt{N}=\{18,20, 22, 24, 26, 28, 30 \}$.  The solid circles
  are the results of the simulations. The dashed line is just to guide the eye. The dash-dotted line is
  plotted at the universal value of the cumulant $U^{*}=0.61069$. }
  \label{f:cumulant}
\end{figure}

The $n$-th moment $\langle m^n \rangle$ can be easily calculated from the distribution of the order parameter $P_N(m)$:
\begin{equation}
\langle m^n \rangle = \frac{\sum_{m} m^n P_N(m)}{\sum_{m}  P_N(m)}\,.
\end{equation}
The apparent critical packing fractions $\eta_\text{c}^*(N)$ satisfy the equation $U_N(\eta_\text{c}^*)=U^*$,
and have been read off from  the interpolated line for
the value of $U^{*}=0.61069$~\cite{kamieniarz:93:0}. In figure~\ref{f:cumulant}, we present the cumulants,
$U_N(m)$, as  functions of the packing fraction for different system sizes $N$.
The curves cross at the values of the cumulant higher than the universal value $U^{*}$. Similar behavior was
observed in references~\cite{buhot:05:0,almarza:15:0}.

\begin{figure}[!t]
  \begin{center}
\includegraphics[width=0.50\textwidth,angle=0]{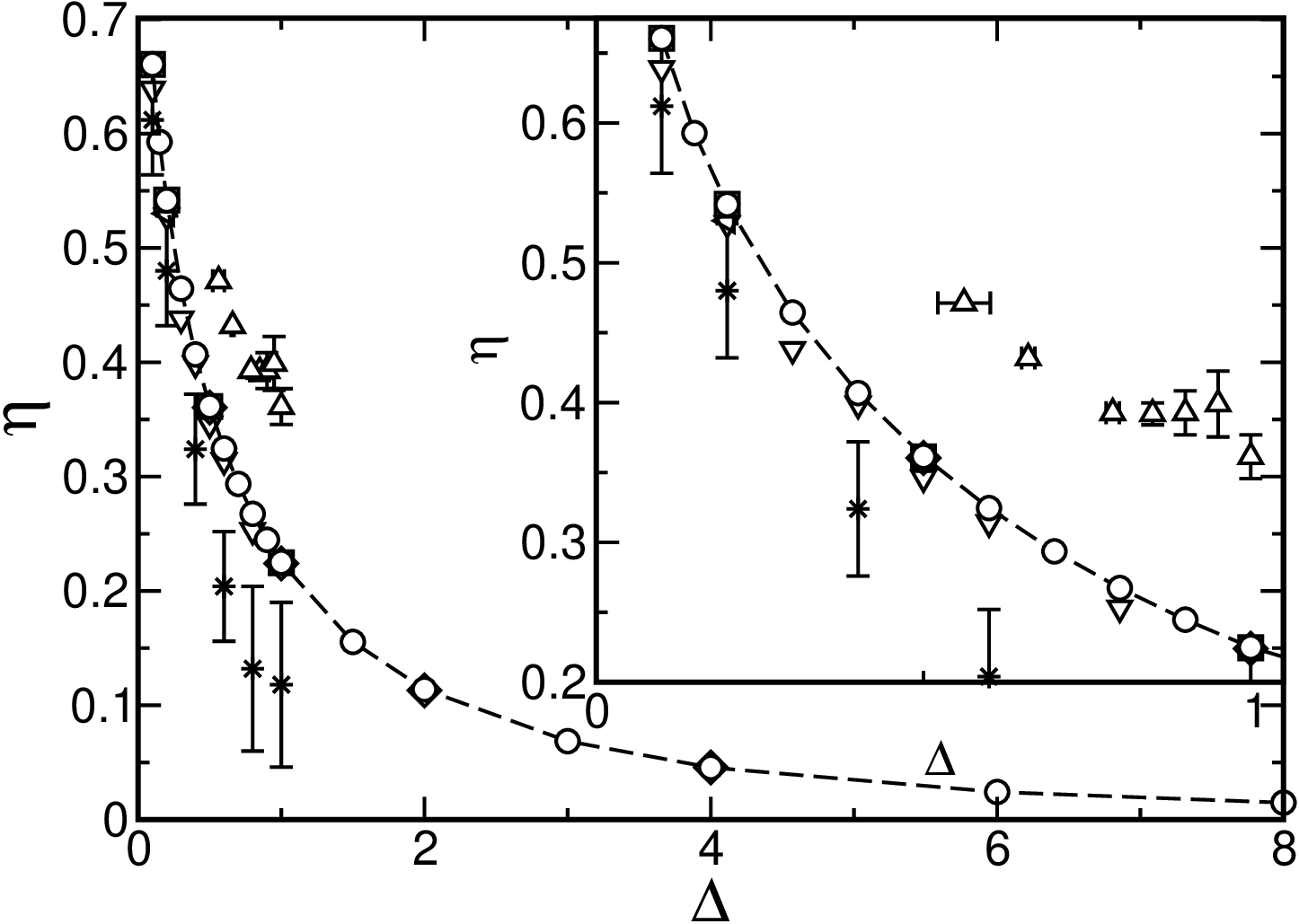}
  \end{center}
  \vspace{-2mm}
  \caption{The critical packing fraction as a function of the non-additivity parameter $\Delta$.
  The open circles show the calculations from this work, the triangles up~--- reference~\cite{nielaba:96:0}, the stars~--- \cite{saija:02:0}, the triangles down~--- \cite{fiumara:14:0}, the squares~--- \cite{almarza:15:0}, the diamonds~--- \cite{buhot:05:0}, the triangle right-hand~--- \cite{munoz-salazar:10:0}. The dashed line is just to guide the eye.}
  \label{f:etac-delta}
\end{figure}

In figure~\ref{f:etac-delta} we present the results of our calculation of the critical packing fractions
for a set of the non-additivity parameter $\Delta$, compared with the results already reported in the literature.
The calculations reported in reference~\cite{nielaba:96:0} significantly overestimate and in reference~\cite{saija:02:0}
significantly underestimate the results of other works~\cite{buhot:05:0,munoz-salazar:10:0,fiumara:14:0,almarza:15:0}.
This is most probably the result of  inappropriate  simulation methods employed in those calculations.
It can be easily noticed that the results of our calculations are in very good agreement with the results of
the calculations reported in reference~\cite{buhot:05:0} and reference~\cite{almarza:15:0}. In all these works, the
authors use cluster algorithms in the simulations and the critical packing fractions are calculated for infinite
systems where different kinds of  finite size scaling analysis were used. In reference~\cite{munoz-salazar:10:0},
the critical point packing fraction was obtained using the method of crossing the reduced second moment.
This method is not as precise as  the method of finite size scaling analysis. In reference~\cite{fiumara:14:0},
the critical packing fraction was estimated from the simulation of finite systems of the order of $N=2000$ molecules.
Thus, the values of the critical packing fractions should be lower than the values of the critical packing fractions obtained
for  infinite systems.

In our work, we employ the same cluster algorithm as in reference~\cite{almarza:15:0}. We  use, however,  a different numerical
algorithm to identify the clusters, and different scaling procedure to obtain the critical packing fraction.
In reference~\cite{buhot:05:0}, a different cluster algorithm is used, where the clusters are built by superimposing
two configurations and identifying the clusters by selecting the groups of overlapping molecules in the sense of
hard core interactions. This method works very well for sufficiently large values of the non-additivity parameter
$\Delta$ but is not as good for small $\Delta$. In reference~\cite{buhot:05:0}, calculations were performed for
$\Delta \in \{ 0.5 , 1.0 , 2.0 , 4.0\}$. The cluster algorithm used in reference~\cite{almarza:15:0} and in our
calculations  works very well for any value of the non-additivity parameters. In reference~\cite{almarza:15:0},
calculations were performed for $\Delta \in \{ 0.1 , 0.2 , 0.5 , 1.0   \}$. In this work, we have performed calculations for all the values of the non-additivity parameter $\Delta$ for which the simulation
results  already existed, and we have extended the calculation to additional large and small non-additivities. We have
performed the calculation for
$\Delta \in \{ 0.1, 0.15, 0.2, 0.3, 0.4, 0.5,  0.6, 0.7, 0.8, 0.9, 1.0, 1.5, 2.0, 3.0, 4.0, 6.0, 8.0   \}$.
The critical packing fractions for this set of parameters  are presented in table~\ref{etactable}.
 \begin{table}[!b]
\caption{The critical packing fractions $\eta_\text{c}$ for infinite systems for different values of the non-additivity parameter $\Delta$.\label{etactable}}
 \vspace{2ex}
\begin{center}
\begin{tabular}{c|c|c|c|c|c|c}
\hline
\hline
$\Delta$ & 0.1         & 0.15        & 0.2        & 0.3         & 0.4        & 0.5       \\
\hline
$\eta_\text{c}$ &  0.6607(30)   & 0.5928(20) & 0.5416(30) & 0.4644(30) & 0.4069(30) &  0.3615(40) \\
\hline
\hline
$\Delta$  & 0.6         & 0.7       &  0.8         & 0.9         & 1.0        & 1.5   \\
\hline
$\eta_\text{c}$  &  0.3246(40) & 0.2936(30) &  0.2674(40)  &  0.2447(40) & 0.2250(30) & 0.1554(30) \\
\hline
\hline
$\Delta$  & 2.0        & 3.0         & 4.0         & 6.0         & 8.0       &    \\
\hline
$\eta_\text{c}$  &  0.1140(40) & 0.0686(30)  & 0.0455(30)  & 0.0241(20) & 0.0148(20) &   \\
\hline
\hline
\end{tabular}
\end{center}
\end{table}
In figure~\ref{f:etac-delta}, the results of our calculations are indistinguishable from the results of calculations in references \cite{buhot:05:0} and \cite{almarza:15:0}. An increase of the non-additivity parameter $\Delta$ results in a decrease of the critical packing fraction as shown in  figure~\ref{f:etac-delta}. In the mixture of non-additive hard disks, we have a competition between the entropy of mixing and the excluded volume effects. When the non-additivity parameter is large, the excluded volume effects are strong and the demixing process takes place at a lower packing fraction.  When the non-additivity parameter is small, the entropy of mixing dominates and the demixing takes place at a higher packing fraction.

\section{Summary and conclusions}

We have calculated the critical packing fractions with high accuracy for the non-additive hard disks mixtures
for a wide range of the non-additivity parameter. We have used Monte Carlo simulations with cluster algorithm
which resulted in rejection-free moves. We have used finite size scaling analysis based on the universal
distribution of the order parameter to determine the critical packing fraction for infinite systems.
The results of our calculations agree very well with the results of the previous calculations where the
finite size scaling analysis was used. The proposed simulation method allows for accurate and unambiguous
determination of the critical packing fraction for any value of the non-additivity parameter $\Delta$.
We hope that the results of our calculations will be the reference point for testing  the results of
approximate theories for hard disks systems. They should also allow one to compare the critical properties of the particles adsorbed
at flat or at curved surfaces, and
to determine in this way the effect of the curvature of the
underlying surface on the critical properties of the adsorbed fluid mixture.

\section*{Acknowledgement}
\noindent
The support from NCN grant No $2012/05/B/ST3/03302$ is acknowledged. We would like to  thank Noe Almarza and Pawe{\l} Rogowski for helpful discussions.


\ukrainianpart

\title{Обчислення критичної точки для бінарних сумішей симетричних неадитивних твердих дисків}

\author{В.Т. Гуздзь, А. Цях}

\address{Інститут фізичної хімії Польської академії наук, Варшава, Польща}

\makeukrtitle

\begin{abstract}
Обчислено значення критичних фракцій заповнення для сумішей симетричних неадитивних твердих дисків.
Цікавою рисою даної моделі є факт, що її внутрішня енергія є нульовою і фазові переходи мають ентропійну природу.
Використано кластерний алгоритм для моделювання Монте Карло у напів-великому ансамблі. Застосовано аналіз скінченно-вимірного
скейлінгу для прецизійного обчислення  критичних фракцій заповнення нескінчених систем  для ширшої області параметра неадитивності
порівняно із попередніми дослідженнями.

\keywords фазове співіснування, критична точка, скінченно-вимірний скейлінг, моделювання Монте Карло
\end{abstract}

\end{document}